\documentclass[manuscript]{acmart}
\usepackage{tabularx}
\usepackage[export]{adjustbox}
\usepackage{subcaption}

\AtBeginDocument{%
  }

\begin{document}

\title{Moonwalk: Advancing Gait-Based User Recognition on Wearable Devices with Metric Learning}


\author{Asaf Liberman$^*$}
\email{asaf_liberman@apple.com}
\orcid{0000-0003-3727-9706}

\author{Oron Levy$^*$}
\email{oron_levy@apple.com}

\author{Soroush Shahi}
\email{shahi@apple.com}

\author{Cori Tymoszek Park}
\email{coripark@apple.com}

\author{Mike Ralph}
\email{iMic@apple.com}

\author{Richard Kang}
\email{runchang_kang@apple.com}

\author{Abdelkareem Bedri}
\email{abedri@apple.com}

\author{Gierad Laput}
\email{gierad@apple.com}

\affiliation{%
  \institution{Apple}
  \country{USA}
}

\renewcommand{\shortauthors}{Liberman et al.}


\begin{abstract}
    Personal devices have adopted diverse authentication methods, including biometric recognition and passcodes. In contrast, headphones have limited input mechanisms, depending solely on the authentication of connected devices. We present Moonwalk, a novel method for passive user recognition utilizing the built-in headphone accelerometer. Our approach centers on gait recognition; enabling users to establish their identity simply by walking for a brief interval, despite the sensor’s placement away from the feet. We employ self-supervised metric learning to train a model that yields a highly discriminative representation of a user's 3D acceleration, with no retraining required. We tested our method in a study involving 50 participants, achieving an average F1-score of 92.9\% and equal error rate of 2.3\%. We extend our evaluation by assessing performance under various conditions (\textit{e.g.}, shoe types and surfaces). We discuss the opportunities and challenges these variations introduce and propose new directions for advancing passive authentication for wearable devices.

\end{abstract}

\keywords{recognition, gait, metric learning}


\maketitle

\footnotetext{  \textbf{*} These authors contributed equally to this work}


\section{Introduction}

In recent years, wireless headphones have surged in popularity, propelled by their improved sound quality and unparalleled convenience. As they continue to proliferate, addressing their security and privacy challenges will become more essential. One of the major challenges is user authentication. Unlike smartphones which offer intrinsic biometric authentication, like face recognition, authentication methods for wearable devices are often extrinsic. For example, smartwatches rely on credentials like passcodes, which can be challenging to input given the small form factor. For smaller wearables like headphones, user authentication is even more difficult, since headphones commonly have no touchscreens, and limited sensors to support common biometric modalities. As a result, wireless headphones today rely on their companion device (\textit{e.g.,} phone or laptop) for authentication. This authentication step usually happens when headphones are paired with a device for the first time, after which they connect to the paired device without further authentication. The lack of an inbuilt authentication scheme creates a security gap where an attacker can use a pair of headphones to listen to incoming private communications without the permission of the rightful user. All these challenges underscore the need for more comprehensive and practical authentication methods, which will improve the user experience for wearables in general, and especially for wireless headphones.

Prior work has explored several sensor-based methods for headphone authentication. The microphone has been used to passively capture intrinsic biometrics related to breathing \cite{10.1145/3081333.3081355} or ear physiology \cite{10.1145/3351239,10.1145/3448113} \cite{7820886,8553015}. Voice recognition methods have found success for decades\cite{tandel2020voicesurvey}, and indeed have been implemented in shared smart home devices to control access to private data \cite{applevoice}\cite{googlevoice}. Another sensor broadly available in headphones is the inertial measurement unit (IMU), which includes accelerometers, gyroscopes, and sometimes magnetometers. User recognition using the IMU has been achieved using explicitly performed tapping actions \cite{9367286}, or vibrations through the mandible generated by speech \cite{9546415,9841001}. While largely effective, these authentication methods are mostly active in nature, requiring the user to deliberately speak, tap, or perform breathing gestures. These approaches are also sensitive to noise and longitudinal variation. Additionally, these approaches either require training of the recognition model subsequent to user enrollment, or in the case where training is not required, the feature embedding and distance metric are hand-crafted rather than learned for optimal performance.

\begin{figure}[b]
    \centering
    \includegraphics[width=1\linewidth]{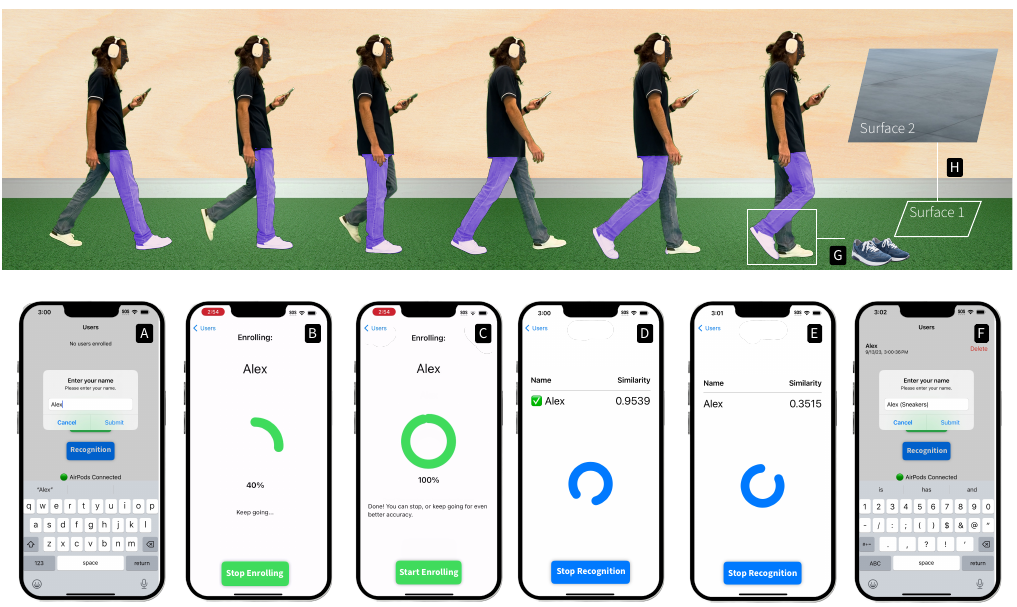}
    \caption{An overview of our user recognition process. Here, a user starts enrolling by entering their name (A), and the enrollment process involves 10 seconds of walking (B). Further, the user can walk for more than 10 seconds to increase the recognition accuracy (C). Our method recognizes the enrolled user based on gait similarity features obtained from headphones accelerometer, requiring no retraining (D). Moreover, our method rejects other people wearing the user's headphones, as noted when a similarity score is low (E). Furthermore, our method supports enrolling different "appearances" (a la facial recognition), such as different shoes (F, G). Finally, our studies show that our method can generalize across different ground surfaces without a need to re-enroll (H).}
    \label{fig:overview}
\end{figure}

Wireless headphones offer freedom of movement, allowing users to freely walk and engage in other activities while wearing them. The passive nature of gait-based recognition means that a wearable can be authenticated using this method without disruption to the user. In this work, we introduce \textbf{Moonwalk}, a gait-based user recognition method using metric learning \cite{metric}. In our approach, we capture gait patterns using a low-power inertial sensor (namely, an accelerometer) found in most wireless headphones and wearable devices. Gait-based recognition is a challenging task because gait patterns can vary with time, shoe type, or floor type. Sensor placement can also impact performance, as the gait's inertial signal diffuses during its propagation from the feet through the body \cite{gafurov2009gait}. To overcome these challenges, we developed a self-supervised metric learning scheme to train our model, making it possible to identify a highly discriminative representation of a user’s 3D acceleration. This metric learning approach has previously been used in face recognition tasks~\cite{kaya2019deep} and has the advantage of not requiring any model retraining when enrolling new users. To the best of our knowledge, we are the first to evaluate the use of metric learning in the context of gait-based user authentication and to demonstrate its strengths over existing methods.

We focused our evaluation on headphones in an effort to address the current gap in authentication capabilities. While headphones are placed at the furthest point from the feet, creating a challenge for gait recognition, gait analysis research shows that meaningful characteristics of the gait are indeed propagated to a headphone\cite{jung2023gaitanalysis}. We evaluated our model on walking data collected from 50 participants. We extended our investigation by evaluating our models on various conditions characterized by different walking surfaces and shoe types. Our results indicate that the accuracy in variable conditions can significantly improve by increasing the size of the model's training set. Furthermore, we present a protocol for enrolling alternate walking styles that comprise a user's natural variation across different environments and shoe types to increase the model's robustness.

Additionally, we leverage the small footprint of our gait recognition model to develop a real-time enrollment and recognition method that runs on-device on a paired smartphone. Figure \ref{fig:overview} shows an iOS app we developed to test our model in real-time. The figure illustrates user enrollment from only 10 seconds of walking data, and testing of the recognition model once enrollment is complete. As part of our evaluation, we assess the usability of this enrollment and recognition process in real time with 19 participants. 

Based on results from our metric learning models, along with results from our real-time evaluations, we believe Moonwalk puts us on the cusp of enabling seamless and passive gait-based authentication on wearable devices.

In summary, the main contributions of this work include:

\begin{enumerate}
\item A gait-based user-recognition model using a self-supervised metric learning scheme. This model was built and evaluated on headphone accelerometer data collected from 50 participants, achieving 92.9\% average \textit{F1-score} and 2.3\% equal error rate (EER) in the user recognition task, using a 10-second window.
\item A real-time enrollment and recognition method that runs on-device, demonstrating the lightweight nature of our approach and the feasibility of practical implementation.
\item Proposed techniques to improve model robustness to unseen conditions that may alter the gait (\textit{e.g.} shoe type, walking surface, fatigue, injury, age).
\item Results from a live evaluation of the usability of our on-device enrollment and recognition method with 19 participants. 
\end{enumerate}

\section{Related Work}

Researchers have extensively explored the concept of passive authentication across various devices, including smartphones~\cite{feng2012continuous, meng2013touch, xu2014towards, jain2015exploring} and wearable devices~\cite{bianchi2016wearable, gafurov2007gait, liu2018vocal}. These devices are now equipped with an array of sensors, such as touchscreens, accelerometers, and gyroscopes, which present opportunities for authenticating users through sensor data. In this section, we narrow our focus to related works that specifically employ sensors positioned on the user's head. Several studies have investigated the use of physiological characteristics for wearable device authentication. These include ear shape~\cite{chen2007human, pflug2012ear}, photoplethysmography (PPG)\cite{zhao2020trueheart}, electrocardiogram\cite{kang2016ecg}, and electroencephalogram~\cite{nakamura2017ear}. For instance, EarEcho~\cite{10.1145/3351239} leverages the unique physical and geometrical characteristics of the ear canal and assesses acoustic features of in-ear sound waves for user authentication. Another approach, EarGait \cite{ferlini2021eargate}, utilizes the microphone of a wearable earpiece to record sounds generated by walking, which propagate through the user's musculoskeletal system. These methods employ audio signals and microphones to capture distinctive user characteristics. However, audio signals captured by microphones are susceptible to environmental noise, which can affect the accuracy and reliability of the authentication process. Additionally, privacy concerns may arise due to the audio data collection process. These limitations highlight the need for alternative approaches. 

Gait analysis has emerged as a promising avenue for user identification. Researchers have explored gait analysis for various applications, including health monitoring and person identification, which holds potential for user authentication~\cite{jarchi2015gait, wan2018survey}. Many studies have focused on gait identification using wearable sensors placed in shoes~\cite{wan2018survey}. Gafurov et al.~\cite{gafurov2009gait} have shown that gait recognition accuracy improves when sensors are positioned closer to the feet, underscoring the challenges associated with gait recognition using head-mounted IMU data. In another approach, Johnston et al. \cite{johnston2015smartwatch} utilized wrist-worn IMU sensors to perform gait recognition using handcrafted features. Similar to our work, some studies have explored gait recognition with sensors positioned on the ear. For example, Jeon et al. \cite{jeon2018biometric} used both accelerometer and gyroscope data from an IMU sensor in a smart earring to identify individuals for an exercise reward system. However, these approaches relied on hand-crafted features or traditional offline classification methods that lack scalability and generalizability for user authentication~\cite{wan2018survey}. 

In contrast to previous work, our research introduces a novel approach by utilizing accelerometer data collected from headphones for gait recognition. Another key differentiator is our utilization of deep metric learning, a technique that has proven highly successful in face recognition~\cite{kaya2019deep}. Our framework allows for a more flexible and user-friendly experience by requiring minimal data for identity enrollment. Moreover, our method demonstrates robustness to noise and different appearances (\textit{e.g.,} varying shoes and surface materials), and lays the foundation for building a generalizable person identification system. 

\section{Method}

\begin{figure}
    \centering
    \includegraphics[width=1\linewidth]{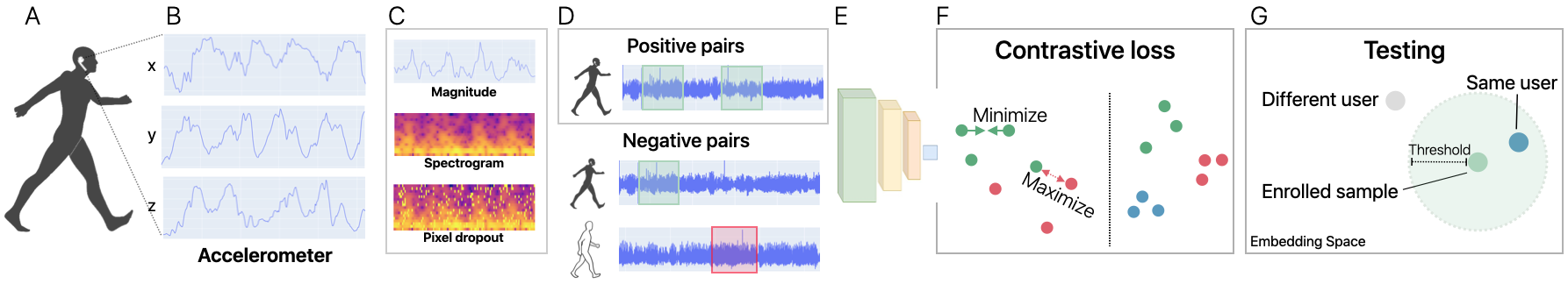}
    \caption{Overview of the Moonwalk pipeline. (a-b) data acquisition from the headphones' accelerometer (3D signal); (c) Pre-processing including conversion to magnitude and then spectrogram, data for training includes the pixel dropout augmentation. (d) In the contrastive training scheme, segments from the same session are defined as positive pairs, and segments from other sessions (belonging to different users) defined as negative pairs. (e) The model outputs embedding for each segment, and (f) trained using a contrastive loss to minimize the distance between embeddings of positive pairs while maximizing the distance between embeddings of negative pairs. The trained model yields a discriminative embeddings space. (g) In the recognition stage, the new sample is compared to an already enrolled user's sample. A distance threshold determines if the sample belongs to that user or not.}
    
    \label{fig:pipeline}
\end{figure}

In this section, we explain the steps we took to build our Moonwalk metric learning model. We also provide details on the datasets used and the metrics we devised for evaluation. Figure \ref{fig:pipeline} provides a full overview of the Moonwalk pipeline.  

\subsection{Datasets}
Our models were built using a generic activity dataset (which we refer to as the \textit{GA dataset}) that we collected from 50 participants (19 females and 31 males). The dataset consists of accelerometer data collected while walking, both indoors and outdoors. Each user was recorded walking in various locations for an average of 10 minutes. The data was collected on six body sensor positions (head, wrist, ankle, chest, shoulder, and thigh) using the built-in accelerometers on AirPods, Apple Watch, and iPhone. Of these, we used the AirPods and ankle (Apple Watch) data (sampled at 100 Hz) to train our models, and the AirPods data for evaluation.

To deepen our understanding of gait-based user recognition challenges and uncover opportunities for advancement in this domain, we collected a second dataset (which we refer to as the \textit{Moonwalk dataset}) focused on walking activities with clear control of adversarial conditions like shoe type and floor surface. The data was collected from 20 participants (5 females, and 15 males) using the AirPods' built-in accelerometer sampled at 100 Hz. Each participant recorded walking sessions with two different pairs of shoes on two different floor surfaces; namely, carpet and concrete. Participants were asked to bring a second set of shoes of their choice. Shoe type was unconstrained. Users primarily wore sneakers, with 81.6\% of all shoes worn being sneakers. Three sets of shoes were flat-soled dress shoes (7.9\%), two were hiking shoes (5.3\%), and one pair each were sandals and high heels (2.6\% each). Of the sneakers, 19 (61.3\%) were flat-soled fashion sneakers, and 12 (31.58\%) were contoured-sole sneakers (\textit{e.g.,} running shoes). When flat- and contoured-sole sneakers are considered as separate types, 6 of the 19 participants had two sets of shoes of the same type, while the remaining 13 had two different types of shoes. Walking data was collected in each of the four combinations (two surfaces and two sets of shoes) for five to ten minutes per combination.

\subsection{Pre-Processing}
\begin{figure}
    \centering
    \includegraphics[width=1\linewidth]{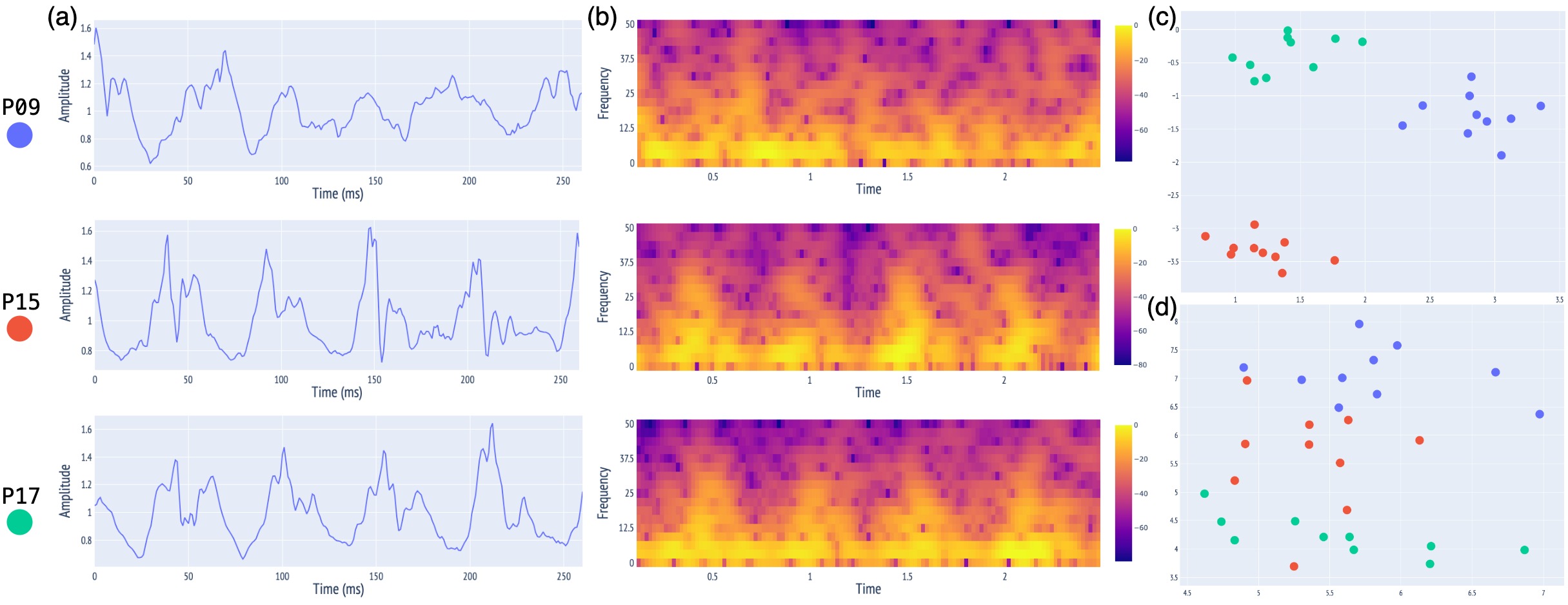}
    \caption{Data representations of several participants. (a) The raw accelerometer signals of Participant 9, Participant 15 and Participant 17. (b) The spectrograms that were calculated from the signals in (a). (c) t-SNE calculated on the embeddings created by the model on the spectrogram input. Participant samples are well separated. (d) t-SNE calculated on features (computed using the raw signal) from the literature \cite{watanabe2020gait}.}
    \label{fig:clustered_spects}
\end{figure}

First, we calculated the single dimensional magnitude of the 3D acceleration signal and converted it into a spectrogram using short-time Fourier transform (STFT). We apply a Hann window for a balanced frequency separation and convert the spectrogram to decibels ($10log_{10}$) for better value range input to the neural network. We use the spectrogram as it encodes both temporal and frequency information inside a 2D array (\textit{i.e.,} an image), which is then easy to feed into a 2D CNN, as described in \cite{Yuwono} and \cite{Jung}. We tested different sample lengths to examine the trade-off between maximizing accuracy and providing a feasible user recognition experience, and we share more details in Section \ref{sec:Experiments}. An example of raw signals with their respective spectrograms, and a distribution of possible embeddings, can be found in Figure \ref{fig:clustered_spects}. To improve model generalizability, we applied a \textit{random pixel dropout} augmentation that zeros random pixels in the spectrogram before feeding it to the model. The pre-processing steps are illustrated in Figure \ref{fig:pipeline}.

\subsection{Training}
We use a self-supervised metric learning scheme \cite{metric} to build a representation of the biometric signature of the user's gait. Figure \ref{fig:pipeline} illustrates the pipeline used for the training, enrollment, and recognition processes. The models were trained on data from the head and ankle sensors (but were enrolled and tested only on the head data). We chose metric learning since it fits well with the task at hand, as it is used to train a model to transfer a signal into an embedding space where distances between similar objects (as defined in training) are closer than distances between dissimilar objects. This means that in the resulting target distribution, it is easier to separate clusters of related samples, which in our case are signals that originate from the same user. This makes metric learning scale well to unfamiliar data, without the need for retraining. For the model, we split a U-Net CNN architecture~\cite{ronneberger2015u} and took the first half as a feature encoder that transforms signals into embedding vectors.



Next, we employ the \emph{NT-Xent} loss, as described in \cite{chen2020simple} and \cite{chen2020big}. This requires contrasting a \textit{positive} pair of samples (\textit{i.e.,} two samples that should belong to the same cluster in the representation space) with \textit{negative} samples (\textit{i.e.}, two samples that should belong to different clusters in the representation space). To achieve this in an unsupervised manner, data augmentation techniques have been used to take a sample and create two transformed versions of it, thus creating a positive pair. Since we have “labeled” data (\textit{i.e.}, we know to which participant, sensor position, surface material and shoe type the samples belong), with long enough periods of activity, we can take two non-overlapping samples of a single participant's activity and use them as a positive pair, having “natural” augmentations that best represent the variance in the signal domain. Since these samples originate from the same walking session, we add augmentations (\textit{i.e.,} pixel dropout) to improve generalization. The resulting model's representation clusters together walking samples from the same user, while distancing them from samples from other users. This enables calculating a threshold to assess a new sample's affiliation with each cluster. Hence, authenticating a user is as simple as calculating the distance to the cluster of representations, acquired during the user's enrollment stage.

\subsection{Loss}
We use the NT-Xent loss function described in \cite{chen2020simple} on a batch of positive and negative samples: 
\[
  \ell_{i,j}^{NT-Xent} = -\log \dfrac{\exp (sim(z_i, z_j)/\tau)}{\sum_{k = 1}^{2N} \mathbf{1}_{[k \neq i]}\exp (sim(z_i, z_k)/\tau)}
\]
where $sim(u, v)$ denote the cosine similarity, $(z_i, z_j)$ are a positive pair, $(z_i, z_k)$ are a negative pair, and $\tau$ denotes the temperature, which regulates the loss. As such, each term in the loss includes a division of the similarity of a positive pair (numerator) by the summed similarities of the first member of the positive pair with all the other (negative) members in the batch (denominator). Back-propagating on this loss encourages a smaller cosine distance in the representation space between the positive pair, and a larger cosine distance between each sample of that pair and each of the negative samples. Thus, the model is trained to output a representation that captures a person's distinct gait. 


\subsection{Testing Procedure and Evaluation Metrics}
In this section, we describe our procedure for understanding how well our model would fare in a user recognition task. Specifically, we conducted a per-user evaluation to determine if, after a short enrollment period, the model can distinguish whether a new entry belongs to that same enrolled user. Here, for each test user, we evenly sampled ten enrollment periods (with a 50\% overlap). We then used the model to extract the pre-processed periods into representation vectors (embeddings), and used their average to create the user's template embedding, which marks the user's cluster center. We randomly sampled 40 “unseen” (\textit{i.e.,} not enrolled on) test samples  and checked whether they fall within a threshold distance from that user's cluster, to get the number of true positives and false negatives. Finally, we performed the same test for embeddings of all other users relative to the enrolled user's template (fifteen from each user), to get metrics on false positives and true negatives. 
Finally, from this, we also obtain a set of confusion matrices and F1-scores. 

To compare the model's performance regardless of the choice of threshold (which would dictate the final user recognition prediction), we iterated over a range of thresholds, calculated the per-user F1-score, averaged along users, and selected the best averaged F1-score. We also calculate the averaged false accept rate (FAR) and false reject rate (FRR) metrics. Finally, we find the equal error rate (EER, the point where both curves meet) from the FAR and FRR curves. These metrics (F1 and EER) are heavily used in literature and allow us to compare our model's performance to other models. Further, they are both threshold invariant, and accommodate biased data well, e.g. where predictions are very different in prevalence, like in our case (there are significantly more negative predictions than positive predictions).
\section{Evaluation and Results}
\label{sec:Experiments}

In this section, we comprehensively describe the sets of experiments we conducted to evaluate our metric learning model, along with detailed results.

\subsection{Comparison with Conventional Methods}

We first start our evaluation by comparing our metric learning approach with conventional methods used for gait-based recognition. To the best of our knowledge, the domain of gait-based recognition does not have established benchmarks. Therefore, we demonstrate the viability of metric learning in this task by comparing its ability to create separable embeddings from data with other choices of feature extraction commonly used in gait-based recognition. Figure \ref{fig:clustered_spects} (d)  shows a 2D t-SNE representation of conventional hand-crafted features as described in \cite{watanabe2020gait}, and Figure \ref{fig:clustered_spects} (c) shows the 2D t-SNE representation for our metric learning embedding space. It is visibly clear from these figures that our approach yields more separable user clusters, which translates to better performance in user recognition tasks.

\subsection{Metric Learning Model}

In this section, we quantify the performance of metric learning models in gait-based user recognition tasks (see Table \ref{tab:results_table} for detailed numerical results). First, we start our evaluation with our GA dataset collected from 50 users. Using \textit{k}-fold cross-validation, we divided 48 out of the 50 participants into 6 independent test sets of 8 users. The test sets are randomly chosen but constant for all the experiments t0, t1, ..., t5. In each experiment we chose a specific test set (for example, t4) to hold out, used the previous test set (\textit{i.e.,} t3) as a validation set and the rest of the participants (the remaining 34) were used as the training set.
After training, we used the best validation score to determine the best model and tested it against the relevant test set (t4). Our evaluation results show that our metric learning approach achieves an average of 92.9\% F1-score and 2.3\% EER in this particular experiment. Figure \ref{fig:gen-clusters} shows a t-SNE projection of the metric learning embedding space demonstrating clear separability between all users.

In addition, we also evaluate model performance given different sample interval lengths. Here, we aim to find a minimally viable duration with which the system can perform sufficiently well. For that, we conducted experiments where we assessed performance with variable windows selected based on our STFT size ranging between 4 to 20 seconds. As shown in Figure \ref{fig:results_vs_duration}, a sample duration of ~10 seconds is at the end of the incline and before the plateau, making it the ideal duration for enrollment and recognition, providing a good balance between accuracy and usability.

\begin{table}
    \centering
    \begin{tabularx}{\textwidth}{|l|l|l|l|X|X|X|X|}
    \hline
    \textbf{\#} & \textbf{Trained On} & \textbf{Enrolled On} & \textbf{Tested On} & \textbf{F1} & \textbf{EER} & \textbf{Alt. Enroll. F1} & \textbf{Alt. Enroll. EER} \\ 
    \hline \hline
    1 & Generic Activity (GA) dataset & GA & GA & \textbf{0.929} & \textbf{0.023} & -- & -- \\
    \hline  
    2 & GA & Surface 1, Shoe 1 & Surface 1, Shoe 1 & 0.895 & 0.05 & -- & -- \\
    \hline
    3 & GA & Surface 1, Shoe 1 & Surface 1, Shoe 2 & 0.471 & 0.129 & 0.868 & 0.078 \\
    \hline
    4 & GA & Surface 1, Shoe 1 & Surface 2, Shoe 1 & 0.845 & 0.055 & -- & -- \\
    \hline \hline
    5 & GA + Moonwalk & Surface 1, Shoe 1 & Surface 1, Shoe 1 & \textbf{0.918} & \textbf{0.034} & -- & -- \\
    \hline
    6 & GA + Moonwalk & Surface 1, Shoe 1 & Surface 1, Shoe 2 & 0.617 & 0.082 & \textbf{0.890} & \textbf{0.057} \\
    \hline
    7 & GA + Moonwalk & Surface 1, Shoe 1 & Surface 2, Shoe 1 & 0.856 & 0.04 & -- & -- \\
    \hline
    \end{tabularx}
      \caption{Evaluation Results. On the first row, the model was trained, enrolled and tested on the \textit{Generic Activity (GA) dataset}. In row 2-4, the model was trained on the GA dataset, and was enrolled and tested on the Moonwalk dataset. In rows 3 and 4, the model was required to generalize between shoe types and surfaces, respectively. Shoe type generalization reduced model performance considerably, while surface generalization still yielded superb performance. In rows 5-7, we added some of our Moonwalk data to model training, which improved generalization, and boosted performance across all test sets. Finally, to mitigate the challenge of generalization across shoe types, we leverage the notion of an 0\textit{alternate appearance} (\textit{i.e.,} enrolling on each shoe separately). Results for enrolling and testing on the alternate shoe enrollment can be seen in the Alternative Enrollment F1 and Alternate Enrollment EER, which suggest that using such a strategy can significantly raise overall accuracy.}
    \label{tab:results_table}
\end{table}


\subsection{Generalizability}

\begin{figure}
    \centering
    \includegraphics[width=1\linewidth]{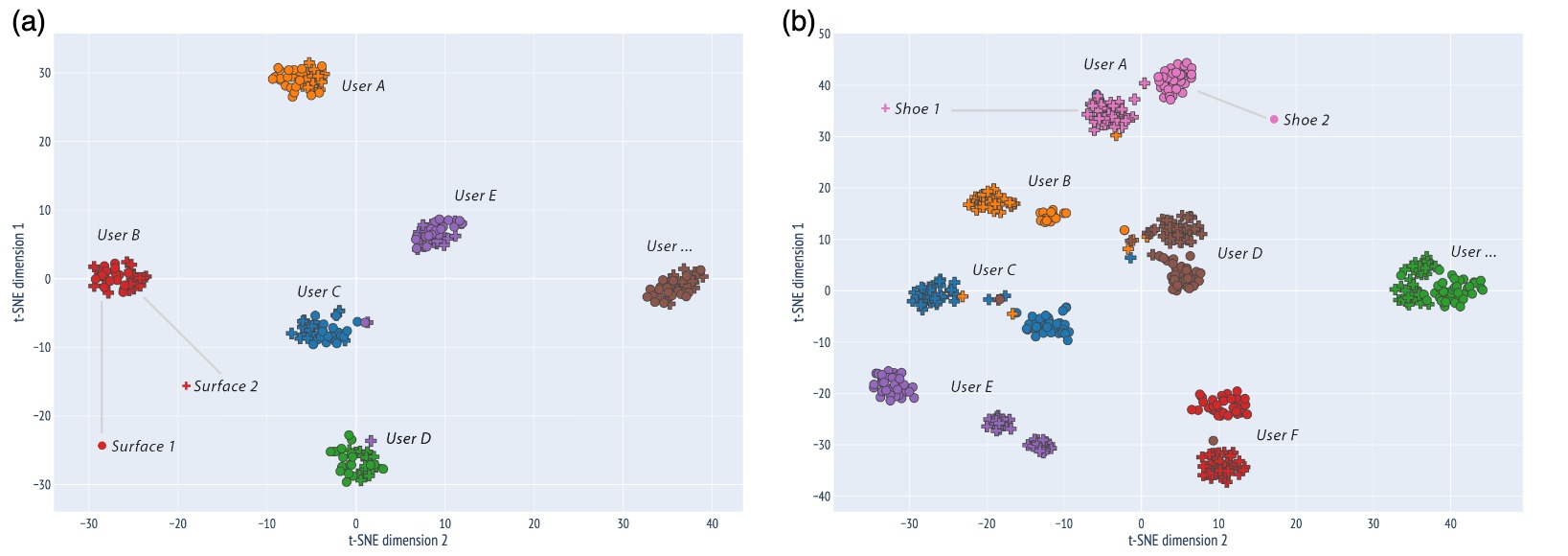}
    \caption{A 2D t-SNE of the distribution of samples in the embedding space. (a) Trained on several floor types (but a single shoe type), and (b) trained on several shoe types. Generalization on surface type is better, as more coherent clusters are created.}
    \label{fig:gen-clusters}
\end{figure}

Encouraged by the results we obtained against our GA dataset, we extend our evaluations to assess the generalizability of this model to the varying conditions captured by the Moonwalk dataset. Table \ref{tab:results_table} shows a breakdown of the results based on how the model was trained, and the enrollment and test conditions. First, we see that the model trained on the GA dataset generalizes reasonably well on different surfaces, achieving 84.5\% average F1-score and 5.5\% EER when enrolling on Surface 1 and testing on Surface 2 with the same shoe (row 4). Figure \ref{fig:gen-clusters}a shows how users' walking data generates unique separable clusters irrespective of surface material.

On the other hand, evaluating generalizability across different footwear resulted in an accuracy drop (row 3). Looking at the metric learning embedding representation (Figure \ref{fig:gen-clusters}b), we see that each user is represented by two not-too-distant clusters. We try to mitigate this by adding more data in the training phase to improve the model's generalizability (row 6). Here we added data by randomly selecting 12 participants from the Moonwalk dataset and adding them to the original GA train-val set (42 participants). This modification seems to greatly improve accuracy (\textit{>14\% jump in F1-score and ~5\% drop in EER}). Compared to prior work, our evaluation uses a greater quantity of data ($n=50$ (GA dataset) + $n=20$ (Moonwalk dataset)), yet these results indicate that we can improve the generalizability results even further by collecting more data.

\subsection{Adaptive Enrollment}
Generalization across different footwear can be alternatively obtained by enrolling the user with different shoes, which is analogous to enrollment with alternate appearances, a commonly-used practice in face recognition tasks (\textit{e.g.,} enrolling with glasses, or a surgical mask). With this approach, there are several templates in the embedding space that are associated with the same user (row 6, \textit{Alt. Enroll}). Using this method, our model generalizes across with a 89.0\% average F1-score.  

We understand that more factors like fatigue, floor incline, and injury can influence the variability in gait patters. To address this challenge we have conducted a preliminary investigation into a concept we refer to as \textit{adaptive enrollment}. Since most current in-ear smart headphones can detect when they are placed in the ear, we are able - by intermittently running the model, as long as the headphones remain in the authenticated user's ear - to identify when an appearance has changed sufficiently to necessitate a new enrollment. This enrolled embedding will join the list of template embeddings that are checked when authenticating that user. To test this idea, we collected data that simulates real world daily use, rich with variability. This data is comprised of different locations, walking paces, surface types and inclines in the same session. The algorithm recognizes a change in conditions, and performs adaptive enrollment. To obtain the most representative template, the algorithm validates the data and selects the most consistent and stable motion segment for the new embedding. Then, it continuously refines the template based on new samples as they accumulate, so it best represents the new appearance.

\begin{figure}
    \centering
    \begin{subfigure}[b]{1.0\textwidth}
        \includegraphics[width=1\linewidth]{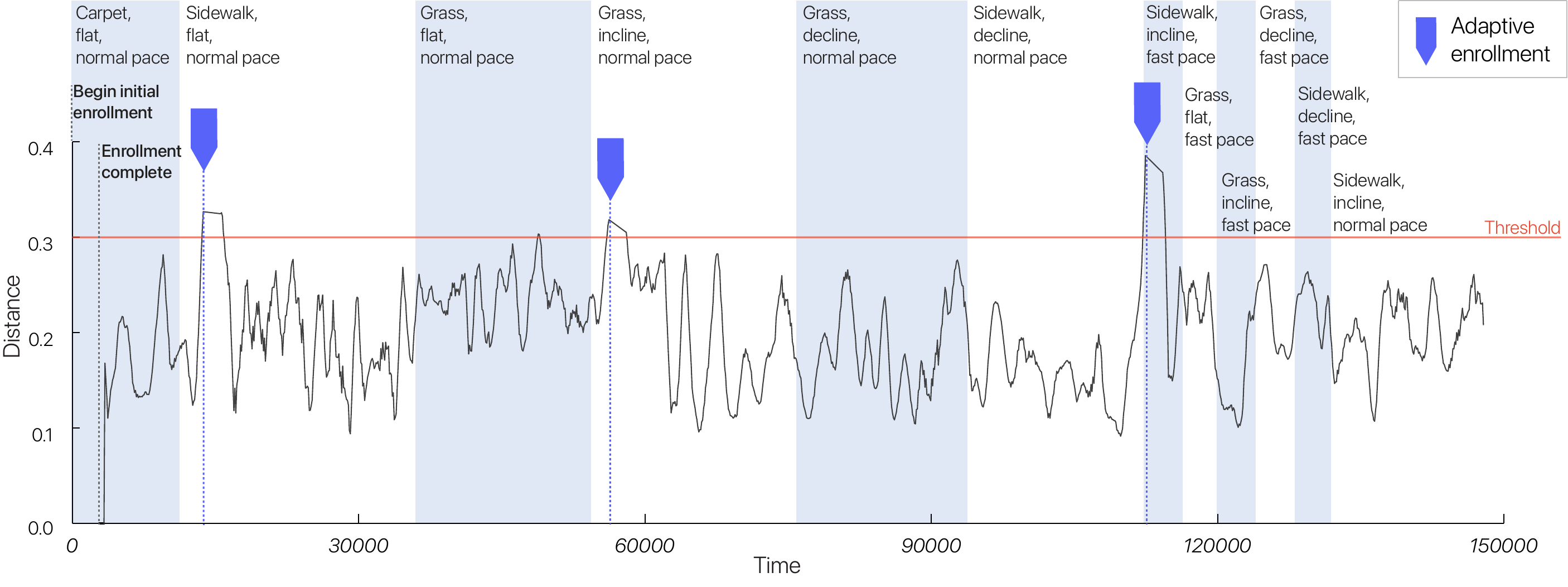}
        \caption{}
        \label{fig:adaptive_lap1}
    \end{subfigure}
    \begin{subfigure}[b]{1.0\textwidth}
        \includegraphics[width=1\linewidth]{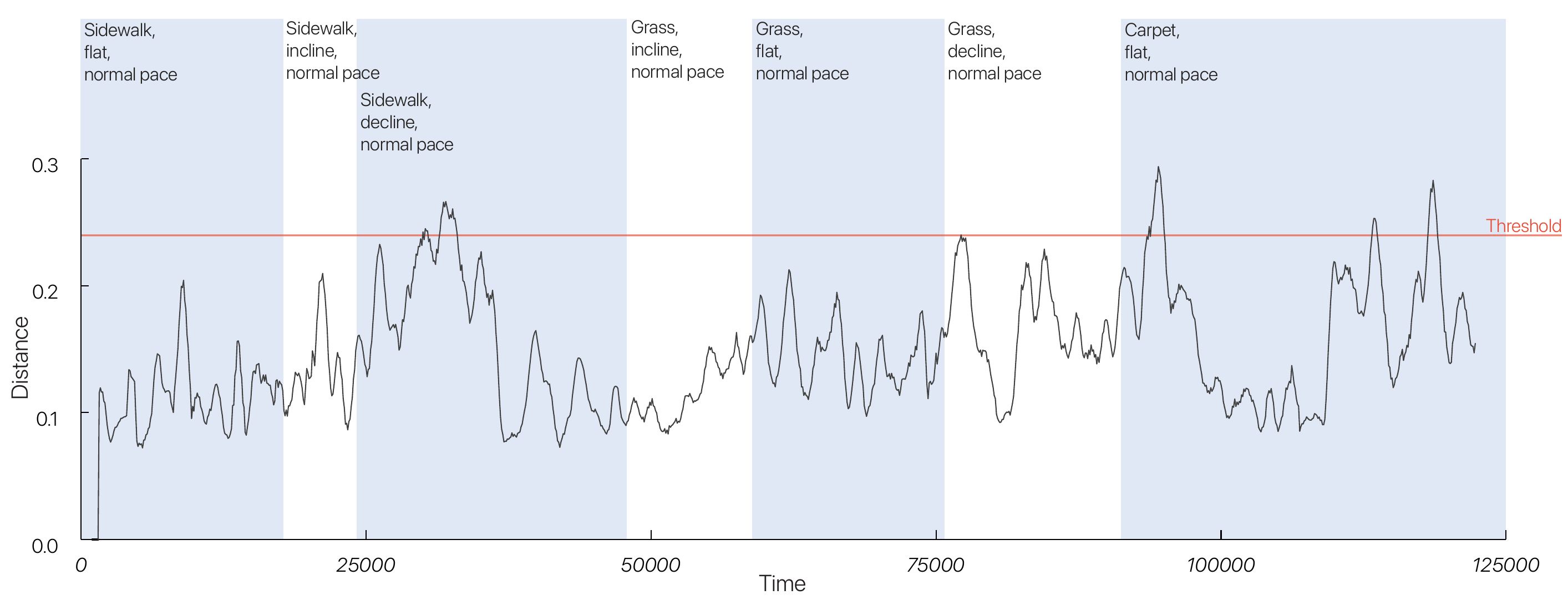}
        \caption{}
        \label{fig:adaptive_lap2}
    \end{subfigure}
    \caption{Illustration of the proposed adaptive enrollment technique in the context of a single user's walking sessions. (a) The enrollment step. The user walks through a variety of conditions, including terrain types, inclinations, and walking paces. The distance from the user's walking embedding is shown over the course of the session. When the distance rises above the threshold (here, 0.3) for multiple time windows, adaptive enrollment occurs, capturing the gait pattern in the problematic segments. Since the headphones are required to remain in-ear for adaptive enrollment to occur, we can assume that the genuine user is still present. As time progresses, the distance approaches the threshold less frequently as more of the user's gait is encompassed within the collected templates. (b) After the adaptive enrollment step, the same set of templates is used to validate the user over a test session. The user again walks through various conditions, and the distance is reported. Here, we do not make the assumption that the headphones remain in-ear, and thus adaptive enrollment does not occur, and we use a stricter threshold of 0.24. The FAR from this user's test session was 7.9\% with 4 total enrolled templates.}
    \label{fig:journey-p111}
\end{figure}

For example, in a single session illustrated in Figure \ref{fig:journey-p111}, a user has walked with normal pace indoors on a level carpet, then outdoors on grass decline, and then walked faster on grass, then moved to a sidewalk, and so on. If our algorithm encounters sufficiently divergent walking data while the headphones remain in the ears, it considers this data as an alternate appearance, and registers a new template when a stable enough sample to be used as a reference is found. For the testing phase, the same user then walked again over similarly varied terrain. The adaptive enrollment templates were used for validation of the subsequent session of that user. We attained a combined FAR of 7.9\% for the 4 enrolled templates (1 from manual enrollment and 3 from adaptive enrollment), and a recall of 94.3\% for the validation session. 

This preliminary investigation highlights the potential of the Adaptive Enrollment method in addressing the generalizability challenge in gait recognition, which unlike face recognition can be influenced by a wide range of factors. In future work, we would like to investigate this method in more naturalistic settings and with a larger pool of participants. Part of that investigation will be to evaluate the trade-off space between the number of enrollments and the risk of false acceptances.  


\begin{figure}
    \includegraphics[width=1\linewidth,left]{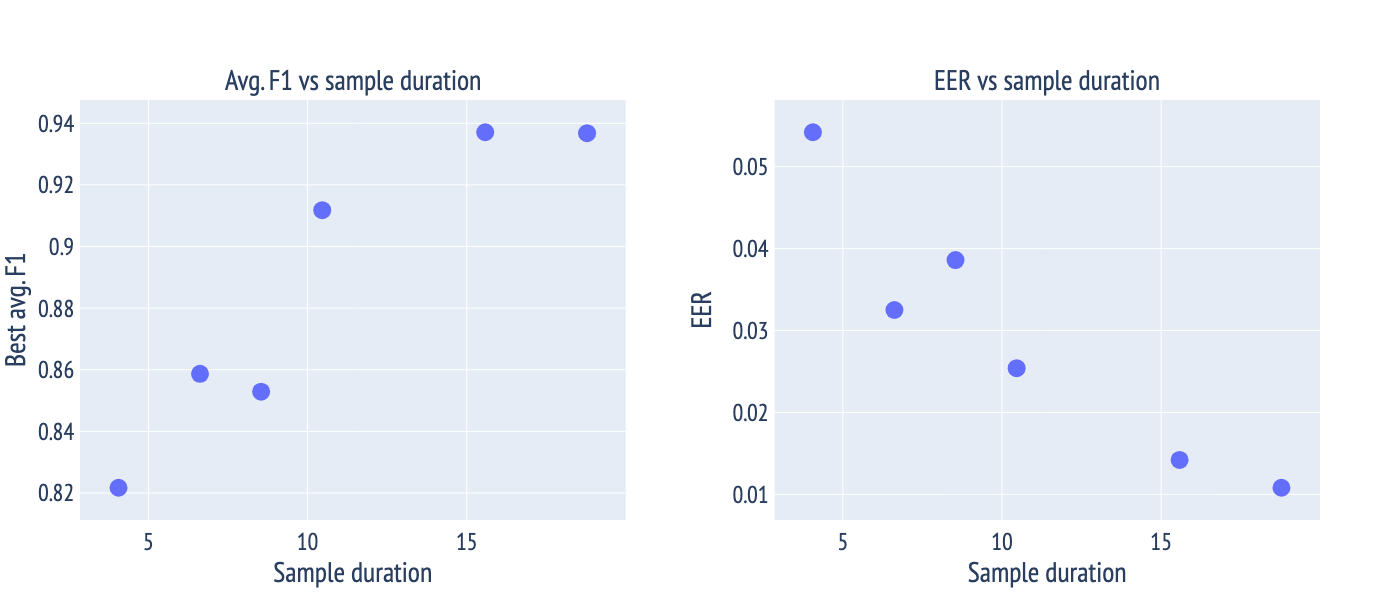}
    \caption{Results for from our GA dataset ($n=50)$ with \textit{k}-fold cross-validation. Performance is highly correlated with sample duration - a longer sample duration improves results, since it contains more temporal context. A plateau can be noticed towards high sample durations, making a sample duration of ~10 ideal.}
    \label{fig:results_vs_duration}
\end{figure}

\subsection{Usability of the Enrollment and Recognition Process}

As part of our evaluation, we developed a process that enables users to enroll their gait information and test the model's accuracy locally on the companion device, \textit{i.e.,} the paired smartphone. This process requires users to provide a minimum of 10 seconds of walking data to enroll, though the app encourages users to continue walking for at least 30 seconds for better performance. We recruited 19 participants to evaluate the usability of the enrollment process after integrating it into an iOS application running on an iPhone 13 Pro. We first gave participants a brief description of how the enrollment and recognition processes worked, and participants were then asked to try the app. At the end of the study, participants filled out NASA-TLX \cite{HART1988139} and SUS \cite{sus} questionnaires to provide feedback on the usability of this interactive process.

Overall, the system received an average overall SUS score of $87.37\pm9.74$ out of 100, indicating that our system has high usability. The NASA-TLX results, rated on a 7-point Likert scale, similarly indicate a positive user experience. Users found the task load to be low in the mental ($1.26\pm0.56$) and temporal ($1.47\pm0.96$) dimensions. Physical load and effort required were rated slightly higher, with average ratings of $2.11\pm1.24$ and $2.05\pm1.47$, respectively, reflecting the inherently physical nature of the walking task. Still, frustration was low ($1.16\pm0.37$), and users felt they were able to complete the task with a high degree of success ($6.47\pm1.39$). These questionnaire results suggest that the required enrollment process would not be perceived as burdensome by users in a practical scenario.

\section{Discussion and Limitations}

In this paper, we show for the first time an extensive evaluation of metric learning in gait-based user recognition tasks using in-ear acceleration signals. In this section, we discuss our main findings, limitations, and areas for future work. 

\subsection{Advantages}

As mentioned, to address the challenges with headphone authentication, we propose employing metric learning methods to build reliable gait recognition models. Our results indicate that within a short period of walking ($\sim$10 seconds), we can verify the identity of the user and reject imposters with a high degree of certainty. This approach has the advantage of being a passive authentication approach, allowing users to establish their identity while walking. Unlike other gait authentication methods, Moonwalk does not require model retraining or fine-tuning to enroll a new user \cite{wan2018survey}. This was only possible by pre-training our metric learning model with a large dataset (GA, $n=50)$, which enabled us to identify an embedding space where each user would have a unique representation of their 3D acceleration data. 
Furthermore, since F1 is the harmonic mean of recall and precision, a high F1 (e.g. like 92.9\%, which was attained in the GA dataset) indicates that both metrics have a high value. This means that the system would need a small number of attempts (often just one attempt) to recognize a user, and that that there would be few false positives (or false accepts). The trade off between these metrics can be adjusted by changing the threshold on the model's confidence score, enabling custom fitting the algorithm per product.

We found several factors that contributed to better models. Firstly, there is a clear correlation between longer sample periods and better performance, as can be seen in Figure \ref{fig:results_vs_duration}. This may be intuitive since longer samples contain more information, which helps the model's prediction. Among all the values we tested, a window of 10 seconds seems to provide the best compromise between accuracy and usability. We also found that taking the magnitude of the 3D signal yields better results, possibly because it reduces noise and normalizes the signal to mitigate the influence of different in-ear headphone positions and different head orientations. We realize that 10s might be too long in some practical scenarios, if, for example, a user wants to hear incoming messages immediately after inserting the headphones. In this scenario, we might defer to the smartphone’s unlocking mechanism to validate that the rightful user is indeed in possession of the devices.

In comparison with previously proposed authentication methods on headphones, our method also holds several advantages. Firstly, while most proposed methods in the literature require internal microphones to sample in-ear audio, our method does not require additional hardware or power-demanding sensors, and it can be implemented in off-the-shelf headphones using built-in low power sensors~\cite{7820886,8553015}. Secondly, our method can also operate under noisy conditions, as opposed to methods that record audio from the user (such as breath sounds)~\cite{10.1145/3081333.3081355}. Lastly, unlike methods that require users to tap their headphones~\cite{9367286} or utter a short phrase to measure the mandible vibrations~\cite{9546415,9841001}, our approach is totally passive and does not require any direct input from the user.

\subsubsection{Comparison to Literature}
It is difficult to directly compare this work to other works in the field. Firstly, few works attempt to characterize gait using an ear-mounted accelerometer, since - as we have also found - positioning the sensor farther from the feet makes recognition using gait harder. Further, in this work we have performed user recognition without constraining the user to surface type or shoe types, over several sessions - which also poses a significant generalization challenge. Under these circumstances, the model achieved an EER of 2.3\% on the GA dataset, and 3.4\% on the Moonwalk dataset. These results are on par or better than others in literature, while still portraying a realistic scenario \cite{gafurov2007gait}, \cite{gafurov2009gait}. For example, in the survey by Wan et al. \cite{wan2018survey}, EER values were reported for 11 accelerometer-based approaches. The average EER across these works was 11.75\%, and all but one of these used a sensor placed at or below the waist. The best-performing approach used an ankle sensor and achieved an EER of 1.6\%, and the second-best performing approach used a waist-located sensor and achieved an EER of 5.6\%. The approach with the highest sensor placement (in the breast pocket) achieved an EER of 14.8\%. Our work performs better than all but the ankle-based approach, while also using a form factor that many users may already own and use. This high-level comparison highlights the strength and reliability of our metric learning approach in varying conditions.

\subsection{Limitations and Future Work}


As a method for authentication, gait recognition presents several challenges. Firstly, while enabling a passive user authentication approach, this method requires instances of active walking, and would not work if the user is idle. We therefore believe that gait recognition can be incorporated as part of a wider user authentication policy, contributing a convenient passive identification method with high uniqueness between users. Second, wheelchair users and people with movement differences may not be able to take advantage of a model pre-trained as we did on typical walking data (\textit{i.e., only on walking data}). However, the metric learning approach we have laid out may be able to distinguish between the uniqueness of users' cane usage, or body motion when manually rolling a wheelchair. Further research could reveal how generalizable our approach is to movement differences. Additionally, our gait recognition model requires a minimum of 10 seconds to achieve good recognition performance, and this limitation may influence the user experience. We believe improving the model such that it requires a shorter time for user recognition is important to increase practical applicability of this method. Another possible direction would be to correlate the gait signal with other wearable devices (such as a smartwatch). A combined signal as an input to the model may reduce the duration required for authentication.

\section{Conclusions}

In this paper, we use a self-supervised metric learning scheme to train a model to distinguish users by their gait pattern using only the built-in accelerometer of wireless headphones. We show that the metric learning approach performs well in both familiar (\textit{i.e.,} previously seen by the model) and unfamiliar conditions. In an evaluation involving 50 participants, we obtain an average F1-score of 92.9\% and EER of 2.9\%. We further demonstrate our method's ability to adapt to the variations of everyday life, including robust generalization to different walking surfaces at baseline, and improved generalization to alternate shoe types when these are considered as "alternate" gaits for the same user. Our results demonstrate that the use of metric learning pushes the envelope on defining novel passive authentication techniques for wearable devices.

\bibliographystyle{ACM-Reference-Format}
\bibliography{acmart}



\end{document}